\documentclass[aps,prb,preprint,superscriptaddress,longbibliography]{revtex4-1}

\usepackage{graphicx}
\usepackage{amssymb}
\usepackage{amsmath}

\begin{document}

\setcounter{MaxMatrixCols}{11}  

\title{Symmetries, Cluster Synchronization, and Isolated Desynchronization in Complex Networks}

\author{Louis M. Pecora}
  \email[]{louis.pecora@nrl.navy.mil}
  \affiliation{U. S. Naval Research Laboratory}
\author{Francesco Sorrentino}
  \affiliation{Department of Mechanical Engineering, University of New Mexico}
\author{Aaron M. Hagerstrom}
  \affiliation{Department of Physics, University of Maryland}
  \affiliation{Institute for Research in Electronics and Applied Physics, University of Maryland}
\author{Thomas E. Murphy}
  \affiliation{Department of Electrical and Computer Engineering, University of Maryland}
  \affiliation{Institute for Research in Electronics and Applied Physics, University of Maryland}
\author{Rajarshi Roy}
  \affiliation{Department of Physics, University of Maryland}
  \affiliation{Institute for Research in Electronics and Applied Physics, University of Maryland}
  \affiliation{Institute for Physical Science and Technology, University of Maryland}

\date{\today}

\begin{abstract}
Synchronization is of central importance in power distribution, telecommunication, neuronal, and biological networks.  Many networks are observed to produce patterns of synchronized clusters, but it has been difficult to predict these clusters or understand the conditions under which they form, except for in the simplest of networks.  In this article, we shed light on the intimate connection between network symmetry and cluster synchronization.  We introduce general techniques that use network symmetries to reveal the patterns of synchronized clusters and determine the conditions under which they persist.  The connection between symmetry and cluster synchronization is experimentally explored using an electro-optic network.  We experimentally observe and theoretically predict a surprising phenomenon in which some clusters lose synchrony while leaving others synchronized.  The results could guide the design of new power grid systems or lead to new understanding of the dynamical behavior of networks ranging 
from neural to social.
\end{abstract}

\maketitle

Synchronization in complex networks is essential to the proper functioning of a wide variety of natural and engineered systems, ranging from electric power grids to neural networks \cite{Motter2013}.  Global synchronization, in which all nodes evolve in unison, is a well-studied effect, the conditions for which are related to the network structure through the master stability function \cite{Pecora1998}.  Equally important, and perhaps more commonplace, is partial, or cluster-synchronization (CS), in which patterns or sets of synchronized elements emerge \cite{Allefeld2007, Ji2013, Zhou2006}.  Recent work on cluster synchronization has been restricted to networks where the synchronization pattern is induced either by tailoring the network geometry or by the intentional introduction of heterogeneity in the time delays or node dynamics \cite{Do2012, Dahms2012, Fu2013, Kanter2011, Rosin2013, Sorrentino2007, Williams2013, Belykh2008}.  These anecdotal studies illustrate the interesting types of cluster 
synchronization that can occur, and suggest a broader relationship between the network structure and synchronization patterns.  Recent studies have begun to draw a connection between network symmetry and cluster synchronization, although all have considered simple networks where the symmetries are apparent by inspection \cite{D'Huys2008, Nicosia2013, Russo2011}. More in-depth studies have been done involving bifurcation phenomena and synchronization in ring and point-symmetry networks \cite{GolubitskyBOOKII,GolubitskySewartBOOK}. Here we address the more common case where the intrinsic network symmetries are neither intentionally produced nor easily discerned.  We present a comprehensive treatment of cluster synchronization, which uses the tools of computational group theory to reveal the hidden symmetries of networks and predict the patterns of synchronization that can arise.  We use irreducible group representations to find a block-diagonalization of the variational equations that can predict the stability 
of the clusters.  We further establish and observe a generic symmetry-breaking bifurcation termed \emph{isolated desynchronization}, in which one or more clusters lose synchrony while the remaining clusters stay synchronized.  The analytical results are confirmed through experimental measurements in a spatio-temporal electro-optic network.  By statistically analyzing the symmetries of several types of networks, as well as electric power distribution networks, we argue that symmetries, clusters, and isolated desynchronization are commonplace and important in many complex networks.

The general dynamical equations to describe a network of $N$ coupled identical oscillators are
\begin{equation}
  \label{eq:1}
  \dot{\bf x}_i(t)={\bf F}({\bf x}_i(t))+\sigma\sum_j A_{ij} {\bf H}({\bf x}_j), \; i=1,...,N,
\end{equation}
where ${\bf x}_i$ is the $n$-dimensional state vector of the $i$-th oscillator, ${\bf F}$ describes the dynamics of each oscillator, $A$ is a symmetric matrix of 1's and 0's that describes the connectivity of the network, $\sigma$ is the coupling strength, and ${\bf H}$ is the output function of each oscillator.  Eq.~(\ref{eq:1}) can be extended to discrete-time systems or more general coupling schemes\cite{Fink2000}.

\begin{figure}[tbp]
  \includegraphics[scale=1.5]{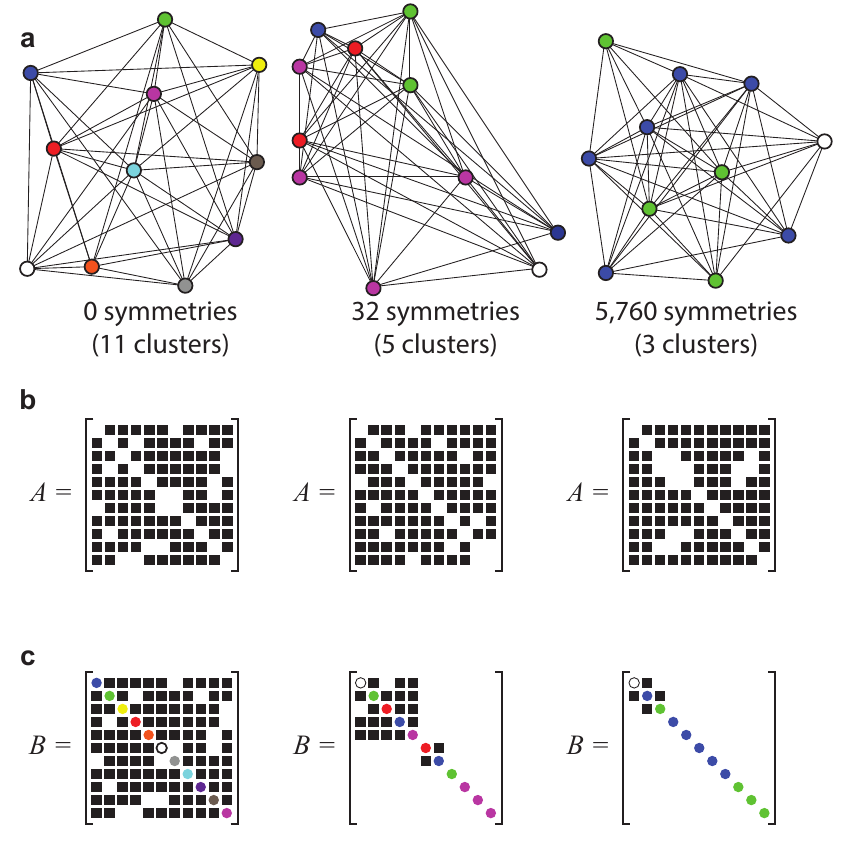}
  \caption{{\bf Three randomly generated networks with varying amounts of symmetry and associated coupling matrices.} (a) Nodes of the same color are in the same synchronization cluster. The colors show the maximal symmetry the network dynamics can have given the graph structure. (b) A graphic showing the structure of the adjacency matrices of each network (black squares are 1, white squares are 0).  (c) Block diagonalization of the coupling matrices $A$ for each network. Colors denote the cluster, as in (a).  The $2 \times 2$ transverse block for the 32 symmetry case comes from one of the IRRs being present in the permutation matrices two times.  The Supplementary Information displays the matrices.
  \label{fig:1}}
\end{figure}

The symmetries of the network form a (mathematical) group ${\cal G}$.  Each symmetry of the group can be described by a permutation matrix $R_g$ that re-orders the nodes in a way that leaves the dynamical equations unchanged (i.e., each $R_g$ commutes with $A$).  The set of symmetries (or automorphisms) \cite{GolubitskyBOOKII,TinkhamBOOK} of a network can be quite large, even for small networks, but they can be calculated from $A$ using widely available discrete algebra routines \cite{Stein,GAP4}.  Figure~\ref{fig:1}a shows three graphs generated by randomly removing 6 edges from an otherwise fully connected 11-node network.  Although the graphs appear similar and exhibit no obvious symmetries, the first instance has no symmetries (other than the identity permutation), while the others have 32 and 5,760 symmetries, respectively. So for even a moderate number of nodes (11) finding the symmetries can become impossible by inspection.

Once the symmetries are identified, the nodes of the network can be partitioned into $M$ clusters by finding the ``orbits'' of the symmetry group: the disjoint sets of nodes that, when all of the symmetry operations are applied, permute among one another \cite{GolubitskyBOOKII}. Because Eq.~\ref{eq:1} is essentially unchanged by the by the permutations the dynamics of the nodes in each cluster can be equal, which is exact synchronization.  Hence, there are M synchronized motions $\{{\bf s}_1,...,{\bf s}_M\}$, one for each cluster.  In Fig.~\ref{fig:1}a, the nodes have been colored to show the clusters.  For the first example, which has no symmetries, the network divides into $M=N$ trivial clusters with one node in each.   The other instances have 5 and 3 clusters, respectively.  Once the clusters are identified, Eq.~(\ref{eq:1}) can be linearized about a state where synchronization is assumed among all of the nodes within each cluster. This linearized equation is the variational equation and it determines 
the stability of the clusters.

Equation (\ref{eq:1}) is expressed in the ``node'' coordinate system, where the subscripts $i$ and $j$ are identified with enumerated nodes of the network.  Beyond identifying the symmetries and clusters, group theory also provides a powerful way to transform the variational equations to a new coordinate system in which the transformed coupling matrix $B = TAT^{-1}$ has a block-diagonal form that matches the cluster structure.  The transformation matrix $T$ is not a simple node re-ordering, nor is it an eigendecomposition of $A$.  The process for computing $T$ is non-trivial, and involves finding the irreducible representations (IRR) of the symmetry group. We call this new coordinate system the ``IRR coordinate system.'' A more detailed description of this process is given later in this article.

Figure ~\ref{fig:1}c shows the coupling matrix $B$ in the IRR coordinate system for the three example networks.  The upper-left block is an $M \times M$ matrix that describes the dynamics within the synchronization manifold.  The remaining diagonal blocks describe motion transverse to this manifold and so are associated with loss of synchronization.  Thus, the diagonalization completely decouples the transverse variations from the synchronization block, and partially decouples the variations among the transverse directions. In this way the stability of the synchronized clusters can be calculated using the separate, simpler, lower dimensional ODEs of the transverse blocks to see if the non-synchronous transverse behavior decays to zero.

The general form of the transformed variational equations for $M$ clusters is,
\begin{equation} \label{eq:2}
\dot{{\pmb \eta}}(t)=\left[\sum_{m=1}^{M} E^{(m)} \otimes D{\bf F}({\bf s}_m(t)) +  \sigma B\otimes I_n \sum_{m=1}^{M} J^{(m)} \otimes D{\bf H}({\bf s}_m(t)) \right] {\pmb \eta}(t),
\end{equation}
where we have linearized about synchronized cluster states $\{{\bf s}_1,...,{\bf s}_M\}$, ${\pmb \eta}(t)$ is the vector of variations of all nodes transformed to the IRR coordinates, $D{\bf F}$ and $D{\bf H}$ are the Jacobians of the nodes' vector field and coupling function, respectively, and $B$ is the block diagonalization of the coupling matrix $A$.  Further details are given in a later section.  We note that this analysis holds for any node dynamics, steady-state, periodic, chaotic, etc.

\begin{figure}[tbp]
  \centering
  \includegraphics[scale=1.5]{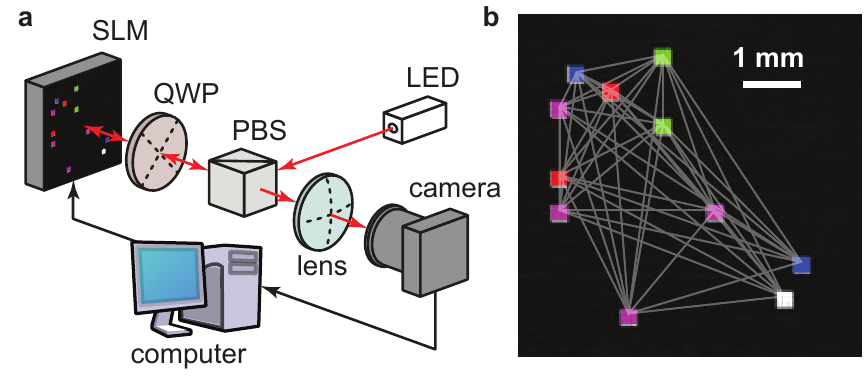}
  \caption{\textbf{Experimental configuration.} a) Light is reflected from the SLM, and passes though polarization optics, so that the intensity of light falling on the camera is modulated according the phase shift introduced by the SLM. Coupling and feedback are implemented by a computer. b) An image of the SLM recorded by the camera in this configuration. Oscillators are shaded to show which cluster they belong to, and the connectivity of the network is indicated by superimposed gray lines.  The phase shifts applied by the square regions are updated according to equation (\ref{eq:3}).
  \label{fig:2}}
\end{figure}

Figure~\ref{fig:2}a shows the optical system used to study cluster synchronization.  Light from a 1550 nm light emitting diode (LED) passes through a polarizing beamsplitter (PBS) and quarter wave plate (QWP), so that it is circularly polarized when it reaches the spatial light modulator (SLM).  The SLM surface imparts a programmable spatially-dependent phase shift $x$ between the polarization components of the reflected signal, which is then imaged, through the polarizer, onto an infrared camera\cite {Hagerstrom2012}.  The relationship between the phase shift $x$ applied by the SLM and the normalized intensity ${\cal I}$ recorded by the camera is ${\cal I}(x)=\left( 1-\cos x \right)/2$.  The resulting image is then fed back through a computer to control the SLM.

The dynamical oscillators that form the network are realized as square patches of pixels selected from a $32 \times 32$ tiling of the SLM array.  Figure~\ref{fig:2}b shows an experimentally measured camera frame captured for one of the 11-node networks considered earlier in Fig.~\ref{fig:1}.  The patches have been falsely colored to show the cluster structure, and the links of the network are overlaid to illustrate the connectivity.  The phase shift of the $i$-th region, $x_i$, is updated iteratively according to:
\begin{equation}
x^{t+1}_i=\left[ \beta {\cal I}(x^{t}_i) + \sigma \sum_j A_{ij} {\cal I}(x^{t}_j) + \delta \right] \text{ mod } 2\pi
\label{eq:3}
\end{equation}
where $\beta$ is the self-feedback strength, and the offset $\delta$ is introduced to suppress the trivial solution $x_i=0$.  Eq.~(\ref{eq:3}) is a discrete-time equivalent of Eq.~(\ref{eq:1}).  Depending on the values of $\beta$, $\sigma$ and $\delta$, Eq.~(\ref{eq:3}) can show constant, periodic or chaotic dynamics.  There are no experimentally-imposed constraints on the adjacency matrix $A_{ij}$, which makes this system an ideal platform to explore synchronization in complex networks.

\begin{figure}[tbp]
  \includegraphics[scale=1.0]{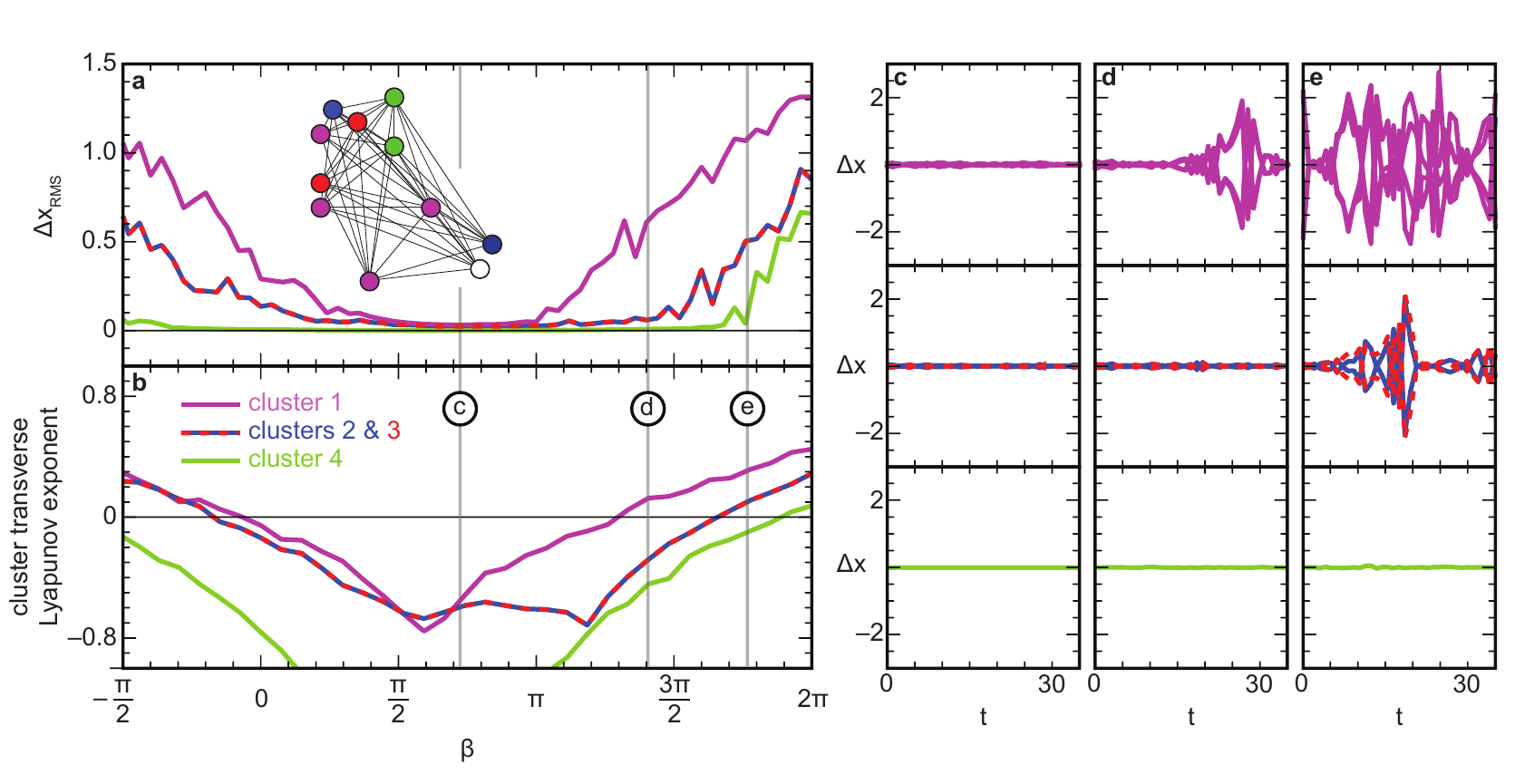}
  \caption{\textbf{Experimental observation of isolated and intertwined desynchronization.} a) Cluster synchronization error as the self-feedback, $\beta$ is varied.  For all cases considered, $\delta = 0.525$ and $\sigma = 0.6\pi$.  Colors indicate the cluster under consideration and are consistent with Fig. \ref{fig:1}. b) MLE calculated from simulation. c-e) Synchronization error time  traces for the four clusters, showing the isolated desychronization of the magenta cluster and the isolated desychronization of the intertwined blue and red clusters.
  \label{fig:3}}
\end{figure}

Figure~\ref{fig:3} plots the time-averaged root-mean square (RMS) synchronization error for all four of the non-trivial clusters shown in Fig.~\ref{fig:2}b, as a function of the feedback strength $\beta$.  The RMS synchronization error was calculated for each cluster as $\Delta x_{\rm RMS} \equiv \left(\overline{\bigl<(x_i^t - \overline{x}^t)^2\bigr>}_T\right)^{1/2}$ where $\left<\bullet\right>_T$ indicates an average over a time interval $T$ (here taken to be 500 iterations) and $\overline\bullet$ denotes a spatial average over the nodes within the cluster.  In Fig.~\ref{fig:3}c-e, we plot the observed intra-cluster deviations $x_i^t - \overline{x}^t$ for three specific values of $\beta$ indicated by the vertical lines in Fig.~\ref{fig:3}a-b, showing different degrees of partial synchronization that can occur, depending on the parameters.

Together, Fig. \ref{fig:3}a and Figs. \ref{fig:3}c-e illustrate two examples of a bifurcation commonly seen in experiment and simulation: isolated desynchronization, where one or more clusters lose stability, while all others remain synchronized. At $\beta=0.72 \pi$ (Fig. \ref{fig:3}c), all four of the clusters synchronize. At $\beta=1.4 \pi$ (Fig. \ref{fig:3}d), the magenta cluster, which contains four nodes, has split into two smaller clusters of 2 nodes each, while the other two clusters remain synchronized.

Between $\beta=0.72 \pi$ and $\beta=1.76 \pi$, two clusters, shown in Fig. \ref{fig:1} as red and blue, undergo isolated desynchronization together. In Fig. \ref{fig:3}a, the synchronization error curves for these two clusters are visually indistinguishable. The synchronization of these two clusters is intertwined: they will always either synchronize together or not at all. While it is not obvious from a visual inspection of the network that the red and blue clusters should form at all, their intertwined synchronization properties can be understood intuitively by examining the connectivity of the network. Each of the two nodes in the blue cluster is coupled to exactly one node in the red cluster. If the blue cluster is not synchronized, the red cluster cannot synchronize because its two nodes are receiving different input. The group analysis treats this automatically and yields a transverse $2 \times 2$ block in Fig.~\ref{fig:1}c.

The isolated desynchronization bifurcations we observe are predicted by computation of the maximum Lyapunov exponent (MLE) of the transverse blocks of Eq.~(\ref{eq:2}), shown in Fig. \ref{fig:3}. The region of stability of each cluster is predicted by a negative MLE. While there are four clusters in this network, there are only three MLEs: the two intertwined clusters are described by a 2-dimensional block in the block-diagonalized coupling matrix $B$. These stability calculations reveal the same bifurcations as seen in experiment.

The existence of isolated desynchronizations in the network experiments raises several questions. Since the network is connected why doesn't the desynchronization pull other clusters out of sync? What is the relation of ID to cluster structure and network symmetry? Is ID a phenomenon that is common to many networks?  We provide answers to all these questions using geometric decomposition of a group which was developed in \cite{MacArthur2009,MacArthur2008}.  This technique enables a finite group to be written as a direct product of subgroups  ${\cal G}={\cal H}_1 \times ...  \times{\cal H}_{\nu}$ where $\nu$ is the number of subgroups and all the elements in one subgroup commute with all the elements in any other subgroup. This means that the set of nodes permuted by one subgroup is disjoint from the set of nodes permuted by any other subgroup. Then each cluster (say, ${\cal C}_j$)  is permuted only by one of the subgroups (say, ${\cal H}_k$), but not by any others. There can be several clusters permuted by 
one subgroup.  This is the case of the red and blue clusters in the 32 symmetry network in Fig.\ref{fig:1}, because the associated ${\cal H}_k$ cannot have a geometric decomposition, but may have a more structured decomposition such as a wreath product \cite{Dionne1996}.

One can show (see the Supplementary Information) that the above decomposition guarantees that the nodes associated with different subgroups all receive the same \emph{total} input from the other subgroups' nodes.  Hence, nodes of each cluster do not see the effects of individual behavior of the other clusters associated with different subgroups. This enables the clusters to have the same synchronized dynamics even when another cluster desynchronizes.  If that state is stable we have ID.

How common is such an ID situation we outlined above?  We have examined statistics for some classes of random and semi-random graph types that suggest that when symmetries are present the opportunity for ID dynamics will be common although the stability for such will depend on the dynamical systems of the network nodes.

We examined 10,000 realizations of three random and semirandom networks: (1) randomly connected nodes (random graphs) similar to Erdos-Renyi graphs \cite{BollobasBOOK}, (2) scalefree tree graphs following Barabasi and Albert \cite{Barabasi1999, Albert2002}, and (3) scalefree using the construction to give predetermined degree distributions characterized by exponent $\gamma$ with same number of connections as in (1) \cite {goh2001universal}. In the Supplementary Information we detail how we generated the realizations and tested for duplicates and statistical relevancy.

\begin{figure}[tbp]
  \includegraphics[width=\textwidth]{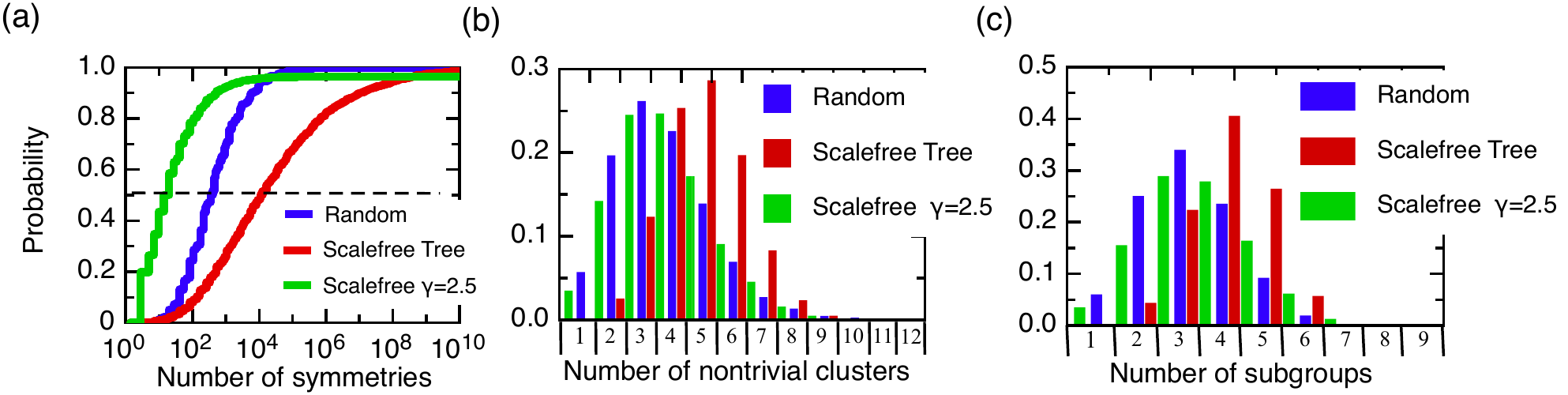}
  \caption{\textbf{ Symmetry, cluster, and subgroup statistics for three types of networks. } The networks are random, Barabasi and Albert (BA in the figure), and the fixed exponent case ($\gamma$ in the figure). The statistics are (a)  the cumulative distribution of the number of symmetries (the dashed line is the median), (b) the counts of the number of nontrivial clusters, and (c) the counts of the number of subgroups in the decomposition.
  \label{fig:4}}
\end{figure}

Figure~\ref{fig:4}a shows the cumulative distribution of symmetries for each type of network.  The $\gamma=2.5$ scalefree graphs generally have fewer symmetries than the other two types.  The Barabasi and Albert scalefree graphs often have many orders of magnitude more symmetries than the others which is a result of their hub and tree structure. All have similar distributions overall, but on different scales of symmetries.  Cases of no symmetry are extremely rare for all graphs in these parameter ranges.  

As shown in Fig.~\ref{fig:4}b,c, almost all graphs for each type have several nontrivial clusters and more than 1 subgroup with the Barabasi and Albert distribution skewing toward somewhat larger numbers.  The median numbers of clusters for the random, Barabasi and Albert, and $\gamma=2.5$ networks are 3, 5, and 3, respectively.  The median numbers of subgroups are 3, 4, and 3, respectively. The percents of cases where the number of subgroups is less than the number of clusters (intertwined cases) are 33\%, 59\%, and 33\%, respectively. Thus, the scenario is present for almost all of these networks to experience ID.

\begin{figure}[tbp]
  \includegraphics[scale=1.5]{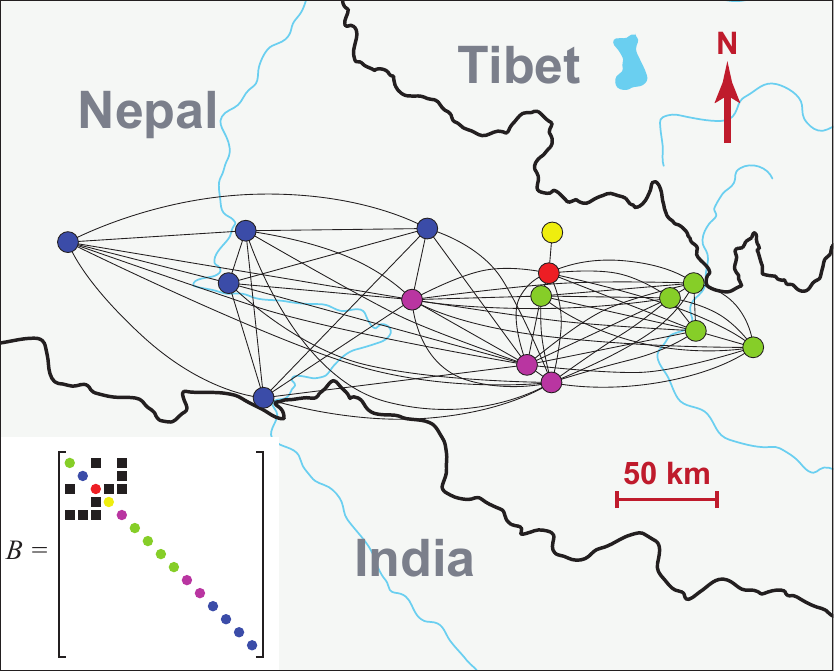}
  \caption{\textbf{Geographical diagram of the Nepal power grid network.}  Colors are used to indicate the computed cluster structure.  The matrix (inset) shows the structure of the diagonalized coupling matrix, analogous to Fig.~\ref{fig:1}a. The diagonal colors indicate which cluster is associated with each column.
  \label{fig:5}}
\end{figure}

Finally, we examined two existing networks: the Nepal power grid \cite{NepalElectReport2011} and the Mesa del Sol electrical grid \cite{abdollahy2012pnm}. We show the Nepal grid since its small size is easier to display in Fig.~\ref{fig:5}. Also shown is the block diagonalization of the coupling matrix. Here we treat the grid analogous to \cite{Motter2013} in which all power stations are identical with the same bidirectional coupling along each edge.  This man-made network has 86,400 symmetries, three nontrivial clusters (plus two trivial ones), and three subgroups (one for each nontrivial cluster). This implies it is possible for this network to split into three sets of synchronized clusters and one of those could lose stability while the others remain synchronized which is ID.  The Mesa del Sol grid has 4096 symmetries, 20 nontrivial clusters, and 10 subgroups.  The network has three intertwined clusters, two with 4 clusters and one with 5 clusters, making ID a possibility.

Many other networks were studied for symmetries in \cite{MacArthur2008, MacArthur2009} for the purpose of finding motifs and redundancies, but not dynamics.  Those networks were Human B Cell Genetic Interactions, C. Elegans genetic interactions, BioGRID data sets (Human, S. cerevisiae Drosphila, and Mus musculus), the internet (Autonomous Systems Level), and the US Power Grid.  All the networks had many symmetries ranging in number from on the order of $10^{13}$ to $10^{11,298}$, and could be decomposed into many subgroups (from 3 to more than 50).  The subgroups were 90\% or more made up of basic factors (not intertwined) consisting of various orders $n$ of the symmetric group $S_n$.  Hence, viewed as dynamical networks, all could show ID in the right situations.

The phenomena of symmetry-induced cluster synchronization and ID appear to be possible in many model, man-made, and natural networks. We've show that ID is explained generally as a manifestation of clusters and subgroup decompositions.  Furthermore, computational group theory can greatly aid in identifying cluster synchronization in complex networks where symmetries are not obvious or far too numerous for visual identification. It also enables explanation of types of desynchronization patterns, and transformation of dynamic equations into more tractable forms. This leads to an encompassing of or overlap with other phenomena which are usually presented as separate.  This list includes (1) remote synchronization \cite{Nicosia2013} in which nodes not directly connected by edges can synchronize (this is just a version of cluster synchronization), (2) some types of chimera states \cite{Abrams2004, Hagerstrom2012} which can appear when the number of trivial clusters is large and the number of nontrivial clusters 
is small, but the clusters are big (see \cite{Laing2009} for some simple examples), (3) partial synchronization where only part of the network is synchronized (shown for some special cases in \cite{Belykh2000}).  We note that although we have concentrated mostly on the maximal symmetry case, we can also examine the  cases of lower symmetry induced by bifurcations that break the original symmetry and the same group theory techniques will apply to those cases. Some of this is developed for simple situations (rings or simple networks with point group symmetry) in Ref. \cite{GolubitskyBOOKII}, but we now have the ability to extend this to arbitrary complex networks. Finally, we note that it is possible to extend this approach to systems with nonidentical oscillators and weighted and/or directional coupling or to hypernetworks \cite{Irving2012}.

\section*{Symmetries, Synchronization Clusters, and Block Diagonalization of the Variational Equations.}

Here we outline the steps necessary to determine the symmetries of the network, obtain the clusters, find the irreducible representations (IRRs), and the most crucial part, calculate the transformation $T$ from the node coordinates to the IRR coordinates that will block-diagonalize $A$, since $A$ commutes with all symmetries of the group \cite{SaganBOOK}.

Using the discrete algebra software it is straightforward to,

(1) Determine the group of symmetries of $A$.

(2) Extract the orbits which give the nodes in each cluster and extract the permutation matrices $R_g$

(3) Using the character table of the group and the traces of the $R_g$'s determine which IRRs are present in the node-space representation of the group. \emph {Remark:} This step is discussed in any book on representations of finite groups (e.g. Ref. \cite{TinkhamBOOK})

(4) Put each $R_g$ into its appropriate conjugacy class.

The next steps are to generate the transformaion $T$ from the group information and they require writing code on top of the discrete algebra software.

(5) For each IRR present construct the projection operator $P^{(l)}$ \cite{TinkhamBOOK} from the node coordinates onto the subspace of that IRR, where $l$ indexes the set of IRRs present.  Thus,
\begin{equation}
P^{(l)}= \frac{d^{(l)}}{h} \sum_{\cal K} \alpha^{(l)}_{\cal K} \sum_{g \in \cal K} R_g
\end{equation}
where $\cal K$ is a conjugacy class, $\alpha^{(l)}_{\cal K}$ is the character of that class for the $l$th IRR, $d^{(l)}$ is the dimension of the $l$th IRR and $h$ is the order (size) of the group. \emph {Remark:} The trivial representation (all IRR matrices=1 and $\alpha^{(l)}=1$) is always present and is associated with the synchronization manifold. All other IRRs are associated with transverse directions.

(6) Use singular value decomposition on $P^{(l)}$ to find the basis for the projection subspace for the $l$th IRR.

(7) Construct $T$ by stacking the row basis vectors of all the IRRs which will form an $N \times N$ matrix.

Once we have $T$ we can transform the variational equations as follows. Let ${\cal C}_m$ be the set of nodes in the $m$th cluster with synchronous motion ${\bf s}_m(t)$. Then the original variational equations about the synchronized solutions are (in vectorial form),

\begin{equation}\label{linv}
\delta \dot{{\bf x}}(t)=\left[\sum_{m=1}^{M} E^{(m)} \otimes D{\bf F}({\bf s}_m(t)) +  \sigma A \sum_{m=1}^{M} E^{(m)} \otimes D{\bf H}({\bf s}_m(t)) \right] \delta {{\bf x}}(t),
\end{equation}
where the $Nn$-dimensional vector $\delta {\bf x}(t)=[\delta {\bf x}_1(t)^T,\delta {\bf x}_2(t)^T,...,\delta {\bf x}_N(t)^T]^T$ and $E^{(m)}$ is an $N$-dimensional diagonal matrix such that
\begin{align}
E^{(m)}_{ii}=\left\{ \begin{array} {ccc} {1,} \quad \mbox{if} \quad {i \in {\mathcal C}_m,} \\ {0,}  \quad \mbox{otherwise,}  \end{array} \right.
\end{align}
$i=1,...,N$.
Note that $\sum_{m=1}^{M} E^{(m)}=I_N$, where $I_N$ is the $N$-dimensional identity matrix.

Applying $T$ to Eq.~(\ref {linv}) we arrive at the variational matrix equation shown in Eq.~(\ref {eq:2}),where ${\pmb \eta}(t)= T \otimes I_n \, \delta {\bf x(t)}$, $J^{(m)}$ is the transformed $E^{(m)}$, and $B$ is the block diagonalization of the coupling matrix $A$. We can write the block diagonal $B$ as a direct sum $\bigoplus_{l=1}^L I_{d^{(l)}} \otimes C^{l}$, where $C^{l}$ is a (generally complex) $p_l \times p_l$ matrix with $p_l=$ the multiplicity of the $l$th IRR in the permutation representation $\{ R_g \}$, $L=$ the number of IRRs present, and $d^{(l)}=$ the dimension of the $l$th IRR, so that $\sum_{l=1}^L d^{(l)} p_l=N$ \cite{Vallintin2009, GoodmanBOOK}. For many transverse blocks $C^{l}$ is a scalar, i.e. $p_l=1$. However, the trivial representation which is associated with the motion in the synchronization manifold has $p_1={M}$. The form of the variational equation for the first examples is shown in Fig.~\ref{fig:1}c. Each block in Fig.~\ref{fig:1}c is governed by a separate variational ODE 
as given in Eq.~(\ref {eq:2}). Note that the vector field ${\bf F}$ can contain a self-feedback term $\beta {\bf x}_i$ as in the experiment and other feedbacks are possible, e.g. row sums of $A_{ij}$, as long as those terms commute with the $R_g$.

{\bf Acknowledgements} We acknowledge help and guidance with computational group theory from Prof. David Joyner of the US Naval Academy and information and computer code from Ben D. MacArthur and Rub\'en J. S\'anchez-Garc\'ia both of the University of Southampton to address the problem of finding group decompositions. Thanks to Vivek Bhandari for providing us a copy of the Nepal Electricity Authority Annual Report \cite{NepalElectReport2011}. Thanks to Shahin Abdollahy and Andrea Mammoli for providing us data on the Mesa Del Sol electric network \cite{abdollahy2012pnm}.

\bibliography{SymsClusterSyncIsoDesync-v7}

\begin{thebibliography}{10}%
\makeatletter
\providecommand \@ifxundefined [1]{%
 \ifx #1\undefined \expandafter \@firstoftwo
 \else \expandafter \@secondoftwo
\fi
}%
\providecommand \@ifnum [1]{%
 \ifnum #1\expandafter \@firstoftwo
 \else \expandafter \@secondoftwo
\fi
}%
\providecommand \enquote [1]{``#1''}%
\providecommand \bibnamefont  [1]{#1}%
\providecommand \bibfnamefont [1]{#1}%
\providecommand \citenamefont [1]{#1}%
\providecommand\href[0]{\@sanitize\@href}%
\providecommand\@href[1]{\endgroup\@@startlink{#1}\endgroup\@@href}%
\providecommand\@@href[1]{#1\@@endlink}%
\providecommand \@sanitize [0]{\begingroup\catcode`\&12\catcode`\#12\relax}%
\@ifxundefined \pdfoutput {\@firstoftwo}{%
 \@ifnum{\z@=\pdfoutput}{\@firstoftwo}{\@secondoftwo}%
}{%
 \providecommand\@@startlink[1]{\leavevmode\special{html:<a href="#1">}}%
 \providecommand\@@endlink[0]{\special{html:</a>}}%
}{%
 \providecommand\@@startlink[1]{%
  \leavevmode
  \pdfstartlink
   attr{/Border[0 0 1 ]/H/I/C[0 1 1]}%
   user{/Subtype/Link/A<</Type/Action/S/URI/URI(#1)>>}%
  \relax
 }%
 \providecommand\@@endlink[0]{\pdfendlink}%
}%
\providecommand \url  [0]{\begingroup\@sanitize \@url }%
\providecommand \@url [1]{\endgroup\@href {#1}{\urlprefix}}%
\providecommand \urlprefix [0]{URL }%
\providecommand \Eprint[0]{\href }%
\@ifxundefined \urlstyle {%
  \providecommand \doi [1]{doi:\discretionary{}{}{}#1}%
}{%
  \providecommand \doi [0]{doi:\discretionary{}{}{}\begingroup
  \urlstyle{rm}\Url }%
}%
\providecommand \doibase [0]{http://dx.doi.org/}%
\providecommand \Doi[1]{\href{\doibase#1}}%
\providecommand \bibAnnote [3]{%
  \BibitemShut{#1}%
  \begin{quotation}\noindent
    \textsc{Key:}\ #2\\\textsc{Annotation:}\ #3%
  \end{quotation}%
}%
\providecommand \bibAnnoteFile [2]{%
  \IfFileExists{#2}{\bibAnnote {#1} {#2} {\input{#2}}}{}%
}%
\providecommand \typeout [0]{\immediate \write \m@ne }%
\providecommand \selectlanguage [0]{\@gobble}%
\providecommand \bibinfo [0]{\@secondoftwo}%
\providecommand \bibfield [0]{\@secondoftwo}%
\providecommand \translation [1]{[#1]}%
\providecommand \BibitemOpen[0]{}%
\providecommand \bibitemStop [0]{}%
\providecommand \bibitemNoStop [0]{.\EOS\space}%
\providecommand \EOS [0]{\spacefactor3000\relax}%
\providecommand \BibitemShut [1]{\csname bibitem#1\endcsname}%
\bibitem{Motter2013}%
  \BibitemOpen
  \bibfield{author}{%
  \bibinfo {author} {\bibfnamefont{A.~E.}\ \bibnamefont{Motter}}, \bibinfo
  {author} {\bibfnamefont{S.~A.}\ \bibnamefont{Myers}}, \bibinfo {author}
  {\bibfnamefont{M.}~\bibnamefont{Anghel}},\ and\ \bibinfo {author}
  {\bibfnamefont{T.}~\bibnamefont{Nishikawa}},\ }%
  \bibfield{title}{%
  \enquote{\bibinfo {title} {Spontaneous synchrony in power-grid networks},}\
  }%
  \bibfield{journal}{%
  \bibinfo {journal} {Nat. Phys.}\ }%
  \textbf{\bibinfo {volume} {9}},\ \bibinfo {pages} {191--197} (\bibinfo {year}
  {2013})%
  \bibAnnoteFile{NoStop}{Motter2013}%
\bibitem{Pecora1998}%
  \BibitemOpen
  \bibfield{author}{%
  \bibinfo {author} {\bibfnamefont{L.M.}\ \bibnamefont{Pecora}}\ and\ \bibinfo
  {author} {\bibfnamefont{T.L.}\ \bibnamefont{Carroll}},\ }%
  \bibfield{title}{%
  \enquote{\bibinfo {title} {Master stability functions for synchronized
  coupled systems},}\ }%
  \bibfield{journal}{%
  \bibinfo {journal} {Physical Review Letters}\ }%
  \textbf{\bibinfo {volume} {80}},\ \bibinfo {pages} {2109--2112} (\bibinfo
  {year} {1998})%
  \bibAnnoteFile{NoStop}{Pecora1998}%
\bibitem{Allefeld2007}%
  \BibitemOpen
  \bibfield{author}{%
  \bibinfo {author} {\bibfnamefont{C.}~\bibnamefont{Allefeld}}, \bibinfo
  {author} {\bibfnamefont{M}~\bibnamefont{M\"uller}},\ and\ \bibinfo {author}
  {\bibfnamefont{J.}~\bibnamefont{Kurths}},\ }%
  \bibfield{title}{%
  \enquote{\bibinfo {title} {Eigenvalue decomposition as a generalized
  synchronizatioun cluster analysis},}\ }%
  \bibfield{journal}{%
  \bibinfo {journal} {Int. J. Bif. Chaos}\ }%
  \textbf{\bibinfo {volume} {17}},\ \bibinfo {pages} {3493--3497} (\bibinfo
  {year} {2007})%
  \bibAnnoteFile{NoStop}{Allefeld2007}%
\bibitem{Ji2013}%
  \BibitemOpen
  \bibfield{author}{%
  \bibinfo {author} {\bibfnamefont{P.}~\bibnamefont{Ji}}, \bibinfo {author}
  {\bibfnamefont{T.K.}\ \bibnamefont{Peron}}, \bibinfo {author}
  {\bibfnamefont{P.J.}\ \bibnamefont{Menck}}, \bibinfo {author}
  {\bibfnamefont{F.A.}\ \bibnamefont{Rodrigues}},\ and\ \bibinfo {author}
  {\bibfnamefont{J.}~\bibnamefont{Kurths}},\ }%
  \bibfield{title}{%
  \enquote{\bibinfo {title} {Cluster explosive synchronization in complex
  networks},}\ }%
  \bibfield{journal}{%
  \bibinfo {journal} {Physical Review Letters}\ }%
  \textbf{\bibinfo {volume} {110}},\ \bibinfo {pages} {218701} (\bibinfo {year}
  {2013})%
  \bibAnnoteFile{NoStop}{Ji2013}%
\bibitem{Zhou2006}%
  \BibitemOpen
  \bibfield{author}{%
  \bibinfo {author} {\bibfnamefont{C.}~\bibnamefont{Zhou}}\ and\ \bibinfo
  {author} {\bibfnamefont{J.}~\bibnamefont{Kurths}},\ }%
  \bibfield{title}{%
  \enquote{\bibinfo {title} {Hierarchical synchronization in complex networks
  with heterogeneous degrees},}\ }%
  \bibfield{journal}{%
  \bibinfo {journal} {CHAOS}\ }%
  \textbf{\bibinfo {volume} {16}},\ \bibinfo {pages} {015104} (\bibinfo {year}
  {2006})%
  \bibAnnoteFile{NoStop}{Zhou2006}%
\bibitem{Do2012}%
  \BibitemOpen
  \bibfield{author}{%
  \bibinfo {author} {\bibnamefont{A-L.Do}}, \bibinfo {author}
  {\bibfnamefont{J.}~\bibnamefont{Hoefener}},\ and\ \bibinfo {author}
  {\bibfnamefont{T.}~\bibnamefont{Gross}},\ }%
  \bibfield{title}{%
  \enquote{\bibinfo {title} {Engineering mesoscale structures with distinct
  dynamical implications},}\ }%
  \bibfield{journal}{%
  \bibinfo {journal} {New Journal of Physics}\ }%
  \textbf{\bibinfo {volume} {14}},\ \bibinfo {pages} {115022} (\bibinfo {year}
  {2012})%
  \bibAnnoteFile{NoStop}{Do2012}%
\bibitem{Dahms2012}%
  \BibitemOpen
  \bibfield{author}{%
  \bibinfo {author} {\bibfnamefont{T.}~\bibnamefont{Dahms}}, \bibinfo {author}
  {\bibfnamefont{J.}~\bibnamefont{Lehnert}},\ and\ \bibinfo {author}
  {\bibfnamefont{E.}~\bibnamefont{Sch\"{o}ll}},\ }%
  \bibfield{title}{%
  \enquote{\bibinfo {title} {Cluster and group synchronization in delay-coupled
  networks},}\ }%
  \bibfield{journal}{%
  \bibinfo {journal} {Physical Review E}\ }%
  \textbf{\bibinfo {volume} {86}},\ \bibinfo {pages} {016202} (\bibinfo {year}
  {2012})%
  \bibAnnoteFile{NoStop}{Dahms2012}%
\bibitem{Fu2013}%
  \BibitemOpen
  \bibfield{author}{%
  \bibinfo {author} {\bibfnamefont{C.}~\bibnamefont{Fu}}, \bibinfo {author}
  {\bibfnamefont{Z.}~\bibnamefont{Deng}}, \bibinfo {author}
  {\bibfnamefont{L.}~\bibnamefont{Huang}},\ and\ \bibinfo {author}
  {\bibfnamefont{X.}~\bibnamefont{Wang}},\ }%
  \bibfield{title}{%
  \enquote{\bibinfo {title} {Topological control of synchronous patterns in
  systems of networked chaotic oscillators},}\ }%
  \bibfield{journal}{%
  \bibinfo {journal} {Physical Review E}\ }%
  \textbf{\bibinfo {volume} {87}},\ \bibinfo {pages} {032909} (\bibinfo {year}
  {2013})%
  \bibAnnoteFile{NoStop}{Fu2013}%
\bibitem{Kanter2011}%
  \BibitemOpen
  \bibfield{author}{%
  \bibinfo {author} {\bibfnamefont{I.}~\bibnamefont{Kanter}}, \bibinfo {author}
  {\bibfnamefont{M.}~\bibnamefont{Zigzag}}, \bibinfo {author}
  {\bibfnamefont{A.}~\bibnamefont{Englert}}, \bibinfo {author}
  {\bibfnamefont{F.}~\bibnamefont{Geissler}},\ and\ \bibinfo {author}
  {\bibfnamefont{W.}~\bibnamefont{Kinzel}},\ }%
  \bibfield{title}{%
  \enquote{\bibinfo {title} {Synchronization of unidirectional time delay
  chaotic networks and the greatest common divisor},}\ }%
  \bibfield{journal}{%
  \bibinfo {journal} {EPL}\ }%
  \textbf{\bibinfo {volume} {93}},\ \bibinfo {pages} {6003--1--6} (\bibinfo
  {year} {2011})%
  \bibAnnoteFile{NoStop}{Kanter2011}%
\bibitem{Rosin2013}%
  \BibitemOpen
  \bibfield{author}{%
  \bibinfo {author} {\bibfnamefont{D.~P.}\ \bibnamefont{Rosin}}, \bibinfo
  {author} {\bibfnamefont{D.}~\bibnamefont{Rontani}}, \bibinfo {author}
  {\bibfnamefont{D.J.}\ \bibnamefont{Gauthier}},\ and\ \bibinfo {author}
  {\bibfnamefont{E.}~\bibnamefont{Sch\"{o}ll}},\ }%
  \bibfield{title}{%
  \enquote{\bibinfo {title} {Control of synchronization patterns in neural-like
  boolean networks},}\ }%
  \bibfield{journal}{%
  \bibinfo {journal} {Physical Review Letters}\ }%
  \textbf{\bibinfo {volume} {110}},\ \bibinfo {pages} {104102} (\bibinfo {year}
  {2013})%
  \bibAnnoteFile{NoStop}{Rosin2013}%
\bibitem{Sorrentino2007}%
  \BibitemOpen
  \bibfield{author}{%
  \bibinfo {author} {\bibfnamefont{F.}~\bibnamefont{Sorrentino}}\ and\ \bibinfo
  {author} {\bibfnamefont{E.}~\bibnamefont{Ott}},\ }%
  \bibfield{title}{%
  \enquote{\bibinfo {title} {Network synchronization of groups},}\ }%
  \bibfield{journal}{%
  \bibinfo {journal} {physical Review E}\ }%
  \textbf{\bibinfo {volume} {76}},\ \bibinfo {pages} {056114} (\bibinfo {year}
  {2007})%
  \bibAnnoteFile{NoStop}{Sorrentino2007}%
\bibitem{Williams2013}%
  \BibitemOpen
  \bibfield{author}{%
  \bibinfo {author} {\bibfnamefont{C.}~\bibnamefont{Williams}}, \bibinfo
  {author} {\bibfnamefont{T.}~\bibnamefont{Murphy}}, \bibinfo {author}
  {\bibfnamefont{R.}~\bibnamefont{Roy}}, \bibinfo {author}
  {\bibfnamefont{F.}~\bibnamefont{Sorrentino}}, \bibinfo {author}
  {\bibfnamefont{T.}~\bibnamefont{Dahms}},\ and\ \bibinfo {author}
  {\bibfnamefont{E.}~\bibnamefont{Sch\"{o}ll}},\ }%
  \bibfield{title}{%
  \enquote{\bibinfo {title} {Experimental observations of group synchrony in a
  system of chaotic optoelectronic oscillators},}\ }%
  \bibfield{journal}{%
  \bibinfo {journal} {Physical Review Letters}\ }%
  \textbf{\bibinfo {volume} {110}},\ \bibinfo {pages} {064104} (\bibinfo {year}
  {2013})%
  \bibAnnoteFile{NoStop}{Williams2013}%
\bibitem{Belykh2008}%
  \BibitemOpen
  \bibfield{author}{%
  \bibinfo {author} {\bibfnamefont{V.}~\bibnamefont{Belykh}}, \bibinfo {author}
  {\bibfnamefont{G.V.}\ \bibnamefont{Osipov}}, \bibinfo {author}
  {\bibfnamefont{V.S.}\ \bibnamefont{Petrov}}, \bibinfo {author}
  {\bibfnamefont{J.K.A.}\ \bibnamefont{Suykens}},\ and\ \bibinfo {author}
  {\bibfnamefont{J.}~\bibnamefont{Vandewalle}},\ }%
  \bibfield{title}{%
  \enquote{\bibinfo {title} {Cluster synchronization in osccillatory
  networks},}\ }%
  \bibfield{journal}{%
  \bibinfo {journal} {CHAOS}\ }%
  \textbf{\bibinfo {volume} {18}},\ \bibinfo {pages} {037106} (\bibinfo {year}
  {2008})%
  \bibAnnoteFile{NoStop}{Belykh2008}%
\bibitem{D'Huys2008}%
  \BibitemOpen
  \bibfield{author}{%
  \bibinfo {author} {\bibfnamefont{O.}~\bibnamefont{D'Huys}}, \bibinfo {author}
  {\bibfnamefont{R.}~\bibnamefont{Vicente}}, \bibinfo {author}
  {\bibfnamefont{T.}~\bibnamefont{Erneux}}, \bibinfo {author}
  {\bibfnamefont{J.}~\bibnamefont{Danckaert}},\ and\ \bibinfo {author}
  {\bibfnamefont{I.}~\bibnamefont{Fischer}},\ }%
  \bibfield{title}{%
  \enquote{\bibinfo {title} {Synchronization properties of network motifs:
  Influence of coupling delay and symmetry},}\ }%
  \bibfield{journal}{%
  \bibinfo {journal} {CHAOS}\ }%
  \textbf{\bibinfo {volume} {18}},\ \bibinfo {pages} {037116} (\bibinfo {year}
  {2008})%
  \bibAnnoteFile{NoStop}{D'Huys2008}%
\bibitem{Nicosia2013}%
  \BibitemOpen
  \bibfield{author}{%
  \bibinfo {author} {\bibfnamefont{V.}~\bibnamefont{Nicosia}}, \bibinfo
  {author} {\bibfnamefont{M.}~\bibnamefont{Valencia}}, \bibinfo {author}
  {\bibfnamefont{M.}~\bibnamefont{Chavez}}, \bibinfo {author}
  {\bibfnamefont{A.}~\bibnamefont{D'az-Guilera}},\ and\ \bibinfo {author}
  {\bibfnamefont{V.}~\bibnamefont{Latora}},\ }%
  \bibfield{title}{%
  \enquote{\bibinfo {title} {Remote synchronization reveals network symmetries
  and functional modules},}\ }%
  \bibfield{journal}{%
  \bibinfo {journal} {Physical Review Letters}\ }%
  \textbf{\bibinfo {volume} {110}},\ \bibinfo {pages} {174102} (\bibinfo {year}
  {2013})%
  \bibAnnoteFile{NoStop}{Nicosia2013}%
\bibitem{Russo2011}%
  \BibitemOpen
  \bibfield{author}{%
  \bibinfo {author} {\bibfnamefont{G.}~\bibnamefont{Russo}}\ and\ \bibinfo
  {author} {\bibfnamefont{J-J.~E.}\ \bibnamefont{Slotine}},\ }%
  \bibfield{title}{%
  \enquote{\bibinfo {title} {Symmetries, stability, and control in nonlinear
  systems and networks},}\ }%
  \bibfield{journal}{%
  \bibinfo {journal} {Physical Review E}\ }%
  \textbf{\bibinfo {volume} {84}},\ \bibinfo {pages} {041929} (\bibinfo {year}
  {2011})%
  \bibAnnoteFile{NoStop}{Russo2011}%
\bibitem{GolubitskyBOOKII}%
  \BibitemOpen
  \bibfield{author}{%
  \bibinfo {author} {\bibfnamefont{M.}~\bibnamefont{Golubitsky}}, \bibinfo
  {author} {\bibfnamefont{I.}~\bibnamefont{Stewart}},\ and\ \bibinfo {author}
  {\bibfnamefont{D.G.}\ \bibnamefont{Schaeffer}},\ }%
  \emph{\bibinfo {title} {Singularities and groups in bifurcation theory}},\
  Vol.~\bibinfo {volume} {II}\ (\bibinfo {publisher} {Springer-Verlag},\
  \bibinfo {address} {New York, NY},\ \bibinfo {year} {1985})%
  \bibAnnoteFile{NoStop}{GolubitskyBOOKII}%
\bibitem{GolubitskySewartBOOK}%
  \BibitemOpen
  \bibfield{author}{%
  \bibinfo {author} {\bibfnamefont{M.}~\bibnamefont{Golubitsky}}\ and\ \bibinfo
  {author} {\bibfnamefont{I.}~\bibnamefont{Stewart}},\ }%
  \emph{\bibinfo {title} {The Symmetry Perspective: From Equilibrium to Chaos
  in Phase Space and Physical Space}},\ Vol.~\bibinfo {volume} {II}\ (\bibinfo
  {publisher} {Berkh\"auser-Verlag},\ \bibinfo {address} {Basel},\ \bibinfo
  {year} {2002})%
  \bibAnnoteFile{NoStop}{GolubitskySewartBOOK}%
\bibitem{Fink2000}%
  \BibitemOpen
  \bibfield{author}{%
  \bibinfo {author} {\bibfnamefont{K.}~\bibnamefont{Fink}}, \bibinfo {author}
  {\bibfnamefont{G.}~\bibnamefont{Johnson}}, \bibinfo {author}
  {\bibfnamefont{D.}~\bibnamefont{Mar}}, \bibinfo {author}
  {\bibfnamefont{T.}~\bibnamefont{Carroll}},\ and\ \bibinfo {author}
  {\bibfnamefont{L.}~\bibnamefont{Pecora}},\ }%
  \bibfield{title}{%
  \enquote{\bibinfo {title} {Three-oscillator systems as universal probes of
  coupled oscillator stability},}\ }%
  \bibfield{journal}{%
  \bibinfo {journal} {Physical Review}\ }%
  \textbf{\bibinfo {volume} {E 61}},\ \bibinfo {pages} {5080--5090} (\bibinfo
  {year} {2000})%
  \bibAnnoteFile{NoStop}{Fink2000}%
\bibitem{TinkhamBOOK}%
  \BibitemOpen
  \bibfield{author}{%
  \bibinfo {author} {\bibfnamefont{M.}~\bibnamefont{Tinkham}},\ }%
  \emph{\bibinfo {title} {Group Theory and Quantum Mechanics}}\ (\bibinfo
  {publisher} {McGraw-Hill},\ \bibinfo {address} {New York, NY},\ \bibinfo
  {year} {1964})\ \bibinfo {note} {the group analysis can be understood with
  only a basic knowledge of finite groups and group representation theory on
  the level of the first three chapters}%
  \bibAnnoteFile{NoStop}{TinkhamBOOK}%
\bibitem{Stein}%
  \BibitemOpen
  \bibfield{author}{%
  \bibinfo {author} {\bibfnamefont{William}\ \bibnamefont{Stein}},\ }%
  \emph{\bibinfo {title} {SAGE: Software for Algebra and Geometry
  Experimentation}}\ (\bibinfo {publisher} {http://www.sagemath.org/sage/,
  http://sage.scipy.org/},\ \bibinfo {year} {2013})%
  \bibAnnoteFile{NoStop}{Stein}%
\bibitem{GAP4}%
  \BibitemOpen
  \bibfield{author}{%
  \bibinfo {author} {\bibfnamefont{The~GAP}\ \bibnamefont{Group}},\ }%
  \emph{\bibinfo {title} {GAP: Groups, Algorithms, and Programming, Version
  4.4}}\ (\bibinfo {publisher} {http://www.gap-system.org},\ \bibinfo {year}
  {2005})%
  \bibAnnoteFile{NoStop}{GAP4}%
\bibitem{Hagerstrom2012}%
  \BibitemOpen
  \bibfield{author}{%
  \bibinfo {author} {\bibfnamefont{A.M.}\ \bibnamefont{Hagerstrom}}, \bibinfo
  {author} {\bibfnamefont{T.E.}\ \bibnamefont{Murphy}}, \bibinfo {author}
  {\bibfnamefont{R.}~\bibnamefont{Roy}}, \bibinfo {author}
  {\bibfnamefont{P.}~\bibnamefont{Hovel}}, \bibinfo {author}
  {\bibfnamefont{I.}~\bibnamefont{Omelchenko}},\ and\ \bibinfo {author}
  {\bibfnamefont{E.}~\bibnamefont{Sch\"{o}ll}},\ }%
  \bibfield{title}{%
  \enquote{\bibinfo {title} {Experimental observation of chimeras in
  coupled-map lattices},}\ }%
  \bibfield{journal}{%
  \bibinfo {journal} {Nature Physics}\ }%
  \textbf{\bibinfo {volume} {8}},\ \bibinfo {pages} {658--61} (\bibinfo {year}
  {2012})%
  \bibAnnoteFile{NoStop}{Hagerstrom2012}%
\bibitem{MacArthur2009}%
  \BibitemOpen
  \bibfield{author}{%
  \bibinfo {author} {\bibfnamefont{B.D.}\ \bibnamefont{MacArthur}}\ and\
  \bibinfo {author} {\bibfnamefont{R.J.}\ \bibnamefont{Sanchez-Garcia}},\ }%
  \bibfield{title}{%
  \enquote{\bibinfo {title} {Spectral characteristics of network redundancy},}\
  }%
  \bibfield{journal}{%
  \bibinfo {journal} {physical Review E}\ }%
  \textbf{\bibinfo {volume} {80}},\ \bibinfo {pages} {026117} (\bibinfo {year}
  {2009})%
  \bibAnnoteFile{NoStop}{MacArthur2009}%
\bibitem{MacArthur2008}%
  \BibitemOpen
  \bibfield{author}{%
  \bibinfo {author} {\bibfnamefont{B.D.}\ \bibnamefont{MacArthur}}, \bibinfo
  {author} {\bibfnamefont{R.J.}\ \bibnamefont{Sanchez-Garcia}},\ and\ \bibinfo
  {author} {\bibfnamefont{J.W.}\ \bibnamefont{Anderson}},\ }%
  \bibfield{title}{%
  \enquote{\bibinfo {title} {On automorphism groups of networks},}\ }%
  \bibfield{journal}{%
  \bibinfo {journal} {Discrete Appl. Math.}\ }%
  \textbf{\bibinfo {volume} {156}},\ \bibinfo {pages} {3525} (\bibinfo {year}
  {2008})%
  \bibAnnoteFile{NoStop}{MacArthur2008}%
\bibitem{Dionne1996}%
  \BibitemOpen
  \bibfield{author}{%
  \bibinfo {author} {\bibfnamefont{B.}~\bibnamefont{Dionne}}, \bibinfo {author}
  {\bibfnamefont{M.}~\bibnamefont{Golubitsky}},\ and\ \bibinfo {author}
  {\bibfnamefont{I.}~\bibnamefont{Stewart}},\ }%
  \bibfield{title}{%
  \enquote{\bibinfo {title} {Coupled cells with internal symmetry: I. wreath
  products},}\ }%
  \bibfield{journal}{%
  \bibinfo {journal} {Nonlinearity}\ }%
  \textbf{\bibinfo {volume} {9}},\ \bibinfo {pages} {559Ð574} (\bibinfo {year}
  {1996})%
  \bibAnnoteFile{NoStop}{Dionne1996}%
\bibitem{BollobasBOOK}%
  \BibitemOpen
  \bibfield{author}{%
  \bibinfo {author} {\bibfnamefont{B.}~\bibnamefont{Bollobas}}\ and\ \bibinfo
  {author} {\bibfnamefont{O.M.}\ \bibnamefont{Riordan}},\ }%
  \emph{\bibinfo {title} {Handbook of Graphs and Networks}}\ (\bibinfo
  {publisher} {Wiley-VCH},\ \bibinfo {address} {Weinheim},\ \bibinfo {year}
  {2003})%
  \bibAnnoteFile{NoStop}{BollobasBOOK}%
\bibitem{Barabasi1999}%
  \BibitemOpen
  \bibfield{author}{%
  \bibinfo {author} {\bibfnamefont{A.-L.}\ \bibnamefont{Barabasi}}\ and\
  \bibinfo {author} {\bibfnamefont{R.}~\bibnamefont{Albert}},\ }%
  \bibfield{title}{%
  \enquote{\bibinfo {title} {Emergence of scaling in random networks},}\ }%
  \bibfield{journal}{%
  \bibinfo {journal} {Science}\ }%
  \textbf{\bibinfo {volume} {286}},\ \bibinfo {pages} {509--512} (\bibinfo
  {year} {1999})%
  \bibAnnoteFile{NoStop}{Barabasi1999}%
\bibitem{Albert2002}%
  \BibitemOpen
  \bibfield{author}{%
  \bibinfo {author} {\bibfnamefont{R.}~\bibnamefont{Albert}}\ and\ \bibinfo
  {author} {\bibfnamefont{A.-L.}\ \bibnamefont{Barabasi}},\ }%
  \bibfield{title}{%
  \enquote{\bibinfo {title} {Statistical mechanics of complex networks},}\ }%
  \bibfield{journal}{%
  \bibinfo {journal} {Reviews of Modern Physics}\ }%
  \textbf{\bibinfo {volume} {74}},\ \bibinfo {pages} {48--96} (\bibinfo {year}
  {2002})%
  \bibAnnoteFile{NoStop}{Albert2002}%
\bibitem{goh2001universal}%
  \BibitemOpen
  \bibfield{author}{%
  \bibinfo {author} {\bibfnamefont{K-I}\ \bibnamefont{Goh}}, \bibinfo {author}
  {\bibfnamefont{B}~\bibnamefont{Kahng}},\ and\ \bibinfo {author}
  {\bibfnamefont{D}~\bibnamefont{Kim}},\ }%
  \bibfield{title}{%
  \enquote{\bibinfo {title} {Universal behavior of load distribution in
  scale-free networks},}\ }%
  \bibfield{journal}{%
  \bibinfo {journal} {Physical Review Letters}\ }%
  \textbf{\bibinfo {volume} {87}},\ \bibinfo {pages} {278701} (\bibinfo {year}
  {2001})%
  \bibAnnoteFile{NoStop}{goh2001universal}%
\bibitem{NepalElectReport2011}%
  \BibitemOpen
  \enquote{\bibinfo {title} {Nepal electricity authority annual report 2011},}\
  \bibinfo {note} {Available at www.nea.org.np}%
  \bibAnnoteFile{NoStop}{NepalElectReport2011}%
\bibitem{abdollahy2012pnm}%
  \BibitemOpen
  \bibfield{author}{%
  \bibinfo {author} {\bibfnamefont{S}~\bibnamefont{Abdollahy}}, \bibinfo
  {author} {\bibfnamefont{O}~\bibnamefont{Lavrova}}, \bibinfo {author}
  {\bibfnamefont{A}~\bibnamefont{Mammoli}}, \bibinfo {author}
  {\bibfnamefont{S}~\bibnamefont{Willard}},\ and\ \bibinfo {author}
  {\bibfnamefont{B}~\bibnamefont{Arellano}},\ }%
  \enquote{\bibinfo {title} {{PNM smart grid demonstration project from
  modeling to demonstration}},}\ in\ \emph{\bibinfo {booktitle} {Innovative
  Smart Grid Technologies (ISGT), 2012 IEEE PES}}\ (\bibinfo {organization}
  {IEEE},\ \bibinfo {year} {2012})\ pp.\ \bibinfo {pages} {1--6}%
  \bibAnnoteFile{NoStop}{abdollahy2012pnm}%
\bibitem{Abrams2004}%
  \BibitemOpen
  \bibfield{author}{%
  \bibinfo {author} {\bibfnamefont{D.M.}\ \bibnamefont{Abrams}}\ and\ \bibinfo
  {author} {\bibfnamefont{S.H.}\ \bibnamefont{Strogatz}},\ }%
  \bibfield{title}{%
  \enquote{\bibinfo {title} {Chimera states for coupled oscillators},}\ }%
  \bibfield{journal}{%
  \bibinfo {journal} {Physical Review Letters}\ }%
  \textbf{\bibinfo {volume} {93}},\ \bibinfo {pages} {174102--1--4} (\bibinfo
  {year} {2004})%
  \bibAnnoteFile{NoStop}{Abrams2004}%
\bibitem{Laing2009}%
  \BibitemOpen
  \bibfield{author}{%
  \bibinfo {author} {\bibfnamefont{C.R.}\ \bibnamefont{Laing}},\ }%
  \bibfield{title}{%
  \enquote{\bibinfo {title} {The dynamics of chimera states in heterogeneous
  kuramoto networks},}\ }%
  \bibfield{journal}{%
  \bibinfo {journal} {Physica D}\ }%
  \textbf{\bibinfo {volume} {238}},\ \bibinfo {pages} {1569--88} (\bibinfo
  {year} {2009})%
  \bibAnnoteFile{NoStop}{Laing2009}%
\bibitem{Belykh2000}%
  \BibitemOpen
  \bibfield{author}{%
  \bibinfo {author} {\bibfnamefont{V.N.}\ \bibnamefont{Belykh}}, \bibinfo
  {author} {\bibfnamefont{I.V.}\ \bibnamefont{Belykh}},\ and\ \bibinfo {author}
  {\bibfnamefont{M.}~\bibnamefont{Hasler}},\ }%
  \bibfield{title}{%
  \enquote{\bibinfo {title} {Hierarchy and stability of partially synchronous
  oscillations of diffusively coupled dynamical systems},}\ }%
  \bibfield{journal}{%
  \bibinfo {journal} {Physical Review E}\ }%
  \textbf{\bibinfo {volume} {62}},\ \bibinfo {pages} {6332--45} (\bibinfo
  {year} {2000})%
  \bibAnnoteFile{NoStop}{Belykh2000}%
\bibitem{Irving2012}%
  \BibitemOpen
  \bibfield{author}{%
  \bibinfo {author} {\bibfnamefont{D.}~\bibnamefont{Irving}}\ and\ \bibinfo
  {author} {\bibfnamefont{F.}~\bibnamefont{Sorrentino}},\ }%
  \bibfield{title}{%
  \enquote{\bibinfo {title} {Synchronization of dynamical hypernetworks:
  dimensionality reduction through simultaneous block-diagonalization of
  matrices},}\ }%
  \bibfield{journal}{%
  \bibinfo {journal} {Physical Review E}\ }%
  \textbf{\bibinfo {volume} {86}},\ \bibinfo {pages} {056102} (\bibinfo {year}
  {2012})%
  \bibAnnoteFile{NoStop}{Irving2012}%
\bibitem{SaganBOOK}%
  \BibitemOpen
  \bibfield{author}{%
  \bibinfo {author} {\bibfnamefont{B.E.}\ \bibnamefont{Sagan}},\ }%
  \emph{\bibinfo {title} {The Symmetric Group}}\ (\bibinfo {publisher}
  {Wadsworth Brooks},\ \bibinfo {address} {Pacific Grove, CA},\ \bibinfo {year}
  {1991})%
  \bibAnnoteFile{NoStop}{SaganBOOK}%
\bibitem{Vallintin2009}%
  \BibitemOpen
  \bibfield{author}{%
  \bibinfo {author} {\bibfnamefont{F.}~\bibnamefont{Vallentin}},\ }%
  \bibfield{title}{%
  \enquote{\bibinfo {title} {Symmetry in semidefinite programs},}\ }%
  \bibfield{journal}{%
  \bibinfo {journal} {Linear Algebra And Its Application}\ }%
  \textbf{\bibinfo {volume} {430}},\ \bibinfo {pages} {360--369} (\bibinfo
  {year} {2009})%
  \bibAnnoteFile{NoStop}{Vallintin2009}%
\bibitem{GoodmanBOOK}%
  \BibitemOpen
  \bibfield{author}{%
  \bibinfo {author} {\bibfnamefont{R.}~\bibnamefont{Goodman}}\ and\ \bibinfo
  {author} {\bibfnamefont{N.R.}\ \bibnamefont{Wallach}},\ }%
  \emph{\bibinfo {title} {Representations and Invariants of the Classical
  Groups}}\ (\bibinfo {publisher} {Cambridge University Press},\ \bibinfo
  {year} {1998})%
  \bibAnnoteFile{NoStop}{GoodmanBOOK}%
\bibitem{Watts1998}%
  \BibitemOpen
  \bibfield{author}{%
  \bibinfo {author} {\bibfnamefont{Duncan~J.}\ \bibnamefont{Watts}}\ and\
  \bibinfo {author} {\bibfnamefont{Steven~H.}\ \bibnamefont{Strogatz}},\ }%
  \bibfield{title}{%
  \enquote{\bibinfo {title} {Collective dynamics of 'smallworld' networks},}\
  }%
  \bibfield{journal}{%
  \bibinfo {journal} {Nature}\ }%
  \textbf{\bibinfo {volume} {393}},\ \bibinfo {pages} {440--442} (\bibinfo
  {year} {1998})%
  \bibAnnoteFile{NoStop}{Watts1998}%
\end{thebibliography}%

\newpage
\section*{Supplementary Information}

{\bf Adjacency, Transformation, and Block-Diagonal Matrices and orbits.} Here are the adjacency matrices, the clusters (group orbits), the transformation, and block-diagonalized coupling matrices for Fig. 1 in more detail.

The 0-symmetry case:

\begin{equation*}
A= \begin{pmatrix}
0&1&1&1&1&1&0&1&1&1&1\\
1&0&1&0&1&1&1&1&0&1&1\\
1&1&0&1&1&1&0&1&1&1&0\\
1&0&1&0&1&1&1&1&1&1&0\\
1&1&1&1&0&1&1&1&1&0&1\\
1&1&1&1&1&0&0&1&1&0&1\\
0&1&0&1&1&0&0&1&1&1&1\\
1&1&1&1&1&1&1&0&1&1&1\\
1&0&1&1&1&1&1&1&0&1&1\\
1&1&1&1&0&0&1&1&1&0&1\\
1&1&0&0&1&1&1&1&1&1&0
   \end{pmatrix}
\end{equation*}
There are no nontrivial clusters, the transformation matrix $T$ is just the identity matrix $I_{11}$, and $B=A$.

The 32-symmetry case:

\begin{equation*}
A= \begin{pmatrix}
0&1&1&1&1&1&0&1&1&1&1\\
1&0&1&1&1&1&0&1&1&1&1\\
1&1&0&1&1&1&1&1&0&1&1\\
1&1&1&0&1&1&1&1&1&1&1\\
0&1&1&1&0&1&1&1&1&1&0\\
1&1&1&1&1&0&1&1&1&1&1\\
1&0&1&1&1&1&0&1&1&1&1\\
1&1&1&1&1&1&1&0&1&0&1\\
1&1&0&1&1&1&1&1&0&1&1\\
1&1&1&1&1&1&1&0&1&0&0\\
1&1&1&1&0&1&1&1&1&0&0
   \end{pmatrix}
\end{equation*}
The nontrivial clusters are the nodes [1, 7], [2, 3, 7, 9], [4, 6], [5, 10] (the numbering of nodes matches the row and column numbers of $A$). The transformation matrix is,

\begin{equation*}
T=  \begin{pmatrix} 0.000&0.000&0.000&0.000&0.000&0.000&0.000&0.000&0.000&1.000&0.000\\
0.000&0.000&-0.707&0.000&-0.707&0.000&0.000&0.000&0.000&0.000&0.000\\
0.000&0.000&0.000&0.707&0.000&0.000&0.000&0.000&0.707&0.000&0.000\\
0.000&0.000&0.000&0.000&0.000&0.000&-0.707&0.000&0.000&0.000&-0.707\\
-0.500&-0.500&0.000&0.000&0.000&-0.500&0.000&-0.500&0.000&0.000&0.000\\
0.000&0.000&0.000&-0.707&0.000&0.000&0.000&0.000&0.707&0.000&0.000\\
0.000&0.000&0.000&0.000&0.000&0.000&-0.707&0.000&0.000&0.000&0.707\\
0.000&0.000&-0.707&0.000&0.707&0.000&0.000&0.000&0.000&0.000&0.000\\
-0.500&0.500&0.000&0.000&0.000&-0.500&0.000&0.500&0.000&0.000&0.000\\
-0.707&0.000&0.000&0.000&0.000&0.707&0.000&0.000&0.000&0.000&0.000\\
0.000&0.707&0.000&0.000&0.000&0.000&0.000&-0.707&0.000&0.000&0.000\\
   \end{pmatrix}
\end{equation*}
And the block diagonal coupling matrix is,
\begin{equation*}
B=  \begin{pmatrix}
0.0&-1.41&0.0&-1.41&-2.00&0.0&0.0&0.0&0.0&0.0&0.0\\
-1.41&1.00&-2.00&2.00&2.83&0.0&0.0&0.0&0.0&0.0&0.0\\
0.0&-2.00&1.00&-1.00&-2.83&0.0&0.0&0.0&0.0&0.0&0.0\\
-1.41&2.00&-1.00&1.00&2.83&0.0&0.0&0.0&0.0&0.0&0.0\\
-2.00&2.83&-2.83&2.83&2.00&0.0&0.0&0.0&0.0&0.0&0.0\\
0.0&0.0&0.0&0.0&0.0&-1.00&1.00&0.0&0.0&0.0&0.0\\
0.0&0.0&0.0&0.0&0.0&1.00&-1.00&0.0&0.0&0.0&0.0\\
0.0&0.0&0.0&0.0&0.0&0.0&0.0&-1.00&0.0&0.0&0.0\\
0.0&0.0&0.0&0.0&0.0&0.0&0.0&0.0&-2.00&0.0&0.0\\
0.0&0.0&0.0&0.0&0.0&0.0&0.0&0.0&0.0&0.0&0.0\\
0.0&0.0&0.0&0.0&0.0&0.0&0.0&0.0&0.0&0.0&0.0\\
   \end{pmatrix}
\end{equation*}

The 5670-symmetry case:

\begin{equation*}
A= \begin{pmatrix} 0&1&1&1&1&1&1&1&1&1&1\\
1&0&1&1&1&1&1&1&1&1&1\\
1&1&0&0&1&1&1&1&1&0&1\\
1&1&0&0&0&1&1&1&0&0&1\\
1&1&1&0&0&1&1&1&0&1&1\\
1&1&1&1&1&0&1&1&1&1&1\\
1&1&1&1&1&1&0&1&1&1&1\\
1&1&1&1&1&1&1&0&1&1&1\\
1&1&1&0&0&1&1&1&0&1&1\\
1&1&0&0&1&1&1&1&1&0&1\\
1&1&1&1&1&1&1&1&1&1&0\\
   \end{pmatrix}
\end{equation*}
The nontrivial clusters are the nodes [2, 6, 1, 7, 8, 11], [3, 5, 9, 10]. The transformation matrix is,
\begin{equation*}
T=  \begin{pmatrix} 0.000&0.000&-1.000&0.000&0.000&0.000&0.000&0.000&0.000&0.000&0.000\\
-0.408&0.000&0.000&0.000&-0.408&-0.408&-0.408&0.000&0.000&-0.408&-0.408\\
0.000&0.500&0.000&0.500&0.000&0.000&0.000&0.500&0.500&0.000&0.000\\
0.000&0.000&0.000&0.000&0.643&-0.448&-0.114&0.000&0.000&0.390&-0.471\\
0.000&0.000&0.000&0.000&-0.522&-0.522&0.224&0.000&0.000&0.596&0.224\\
-0.913&0.000&0.000&0.000&0.183&0.183&0.183&0.000&0.000&0.183&0.183\\
0.000&0.000&0.000&0.000&-0.332&0.568&-0.138&0.000&0.000&0.472&-0.570\\
0.000&0.000&0.000&0.000&-0.066&-0.066&0.847&0.000&0.000&-0.264&-0.451\\
0.000&-0.500&0.000&0.500&0.000&0.000&0.000&0.500&-0.500&0.000&0.000\\
0.000&-0.707&0.000&0.000&0.000&0.000&0.000&0.000&0.707&0.000&0.000\\
0.000&0.000&0.000&-0.707&0.000&0.000&0.000&0.707&0.000&0.000&0.000\\
   \end{pmatrix}
\end{equation*}
And the block diagonal coupling matrix is,
\begin{equation*}
B=  \begin{pmatrix}
0.0&2.45&0.0&0.0&0.0&0.0&0.0&0.0&0.0&0.0&0.0\\
2.45&5.00&-4.90&0.0&0.0&0.0&0.0&0.0&0.0&0.0&0.0\\
0.0&-4.90&2.00&0.0&0.0&0.0&0.0&0.0&0.0&0.0&0.0\\
0.0&0.0&0.0&-1.00&0.0&0.0&0.0&0.0&0.0&0.0&0.0\\
0.0&0.0&0.0&0.0&-1.00&0.0&0.0&0.0&0.0&0.0&0.0\\
0.0&0.0&0.0&0.0&0.0&-1.00&0.0&0.0&0.0&0.0&0.0\\
0.0&0.0&0.0&0.0&0.0&0.0&-1.00&0.0&0.0&0.0&0.0\\
0.0&0.0&0.0&0.0&0.0&0.0&0.0&-1.00&0.0&0.0&0.0\\
0.0&0.0&0.0&0.0&0.0&0.0&0.0&0.0&-2.00&0.0&0.0\\
0.0&0.0&0.0&0.0&0.0&0.0&0.0&0.0&0.0&0.0&0.0\\
0.0&0.0&0.0&0.0&0.0&0.0&0.0&0.0&0.0&0.0&0.0\\
   \end{pmatrix}
\end{equation*}
\\

{\bf Subgroup decomposition and cluster dynamics.} To start let ${\cal H}_k$, a subgroup of ${\cal G}$, permute only cluster ${\cal C}_m$ and $\pi$ be the permutation on the indices of nodes in ${\cal C}_m$ for one permutation $R_g,\; g \in {\cal H}_k$. Assume ${\bf x}_i$ is not in ${\cal C}_m$ so it is not permuted by $R_g$ and recall that $G$ commutes with all permutations in ${\cal G}$, then we have (just concentrating on the terms from ${\cal C}_j$),

\begin{equation*} \label{decompcplg}
\begin{split}
[R_g \dot{\bf x}(t)]_i=\dot{\bf x}_i(t)=...+ \sigma[R_g  A {\bf H}({\bf x})]_i =...+ \sigma[A R_g {\bf H}({\bf x})]_i = ...+ \sigma\sum_{j \in {\cal C}_m} A_{ij} {\bf H}({\bf x}_{\pi(j)}),
\end{split}
\end{equation*}
where $\pi(l)$ is, in general, another node in ${\cal C}_m$ and the sums over other clusters are unchanged. This shows that all nodes in ${\cal C}_m$ are coupled into the $i$th node in the same way (the same $A_{ij}$ factor). Similarly, if we use a permution $R_{g'}$ on the cluster ${\cal C}_{m'}$ containing ${\bf x}_i$ we can show that all the nodes of ${\cal C}_{m'}$ are coupled in the same way to the nodes in ${\cal C}_m$. Hence, nodes of ${\cal C}_{m'}$ each receive the same input sum from the nodes of ${\cal C}_m$ whether the nodes of ${\cal C}_m$ are synchronized or not.  This explains how the cluster ${\cal C}_m$ can become desynchronized, but the nodes of ${\cal C}_{m'}$ can still be synchronized -- they all have the same input despite the ${\cal C}_m$ desynchronization, thus making the ${\cal C}_{m'}$ synchronous state flow invariant. If it is also stable, this is the case of ID. This argument is easily generalized to the case when ${\cal H}_k$ permutes nodes of several clusters as this will just 
add other similar sums to Eq.~(\ref{decompcplg}). The latter case explains the intertwined desynchronization in the experiment and is a more general form of ID.

{\bf Statistics of random graphs.} Random graphs were generated by starting with 25 nodes completely connected and randomly deleting 20 edges.  Scalefree Barabasi and Albert graphs were  based on the original Barabasi and Albert preference algorithm \cite {Barabasi1999} using the \texttt{SAGE} routine \texttt{RandomBarabasiAlbert}. These had 25 nodes with 24 edges and a tree structure. Scalefree graphs with a specific power-law distribution were generated according to \cite {goh2001universal} using $\gamma=2.5, 3.0,$ and $3.5$. 10,000 realizations of each graph type were generated. We tested several 10,000 realizations and we see very little variation in statistics between realizations of the same class leading us to believe that we are sampling fairly and enough to trust our results. We also checked for equivalent (isomorphic) graphs to see how much repetition we had.  The random systems yielded on average 1 equivalent pair per 10000 realizations. The scalefree cases yielded about 5 to 10\% equivalent 
graphs.  Apparently we are not near the maximum number of inequivalent graphs for any of the classes although the results suggest that the scalefree classes are much smaller than the random class. Even with just 100 realizations the main trends in number of symmetries and other statistics are evident although such small samples occasionally miss those symmetry cases that are not too common in the class.

{\bf The scalefree $\gamma$ model.}  The model generates a scale free network with $N$ nodes and $E$ edges and a specified power law degree distribution exponent $\gamma$. Start with $N$ vertices, assign to each vertex $i =1, 2, ...,N$, a weight $w_i = i^{-\mu}$, where the exponent $\mu$ lies in the range [0, 1). Assume that initially no edges are present among the network vertices, then edges are added one by one until $E$ connections are created. For each new edge, two vertices are randomly selected, each one with probability proportional to its weight, and they are connected unless a link already exists or the two selected nodes are the same. By
following this procedure, the resulting network is scale free and the power law degree
distribution $P(k) \sim  k^{-\gamma}$, with exponent $\gamma =(1+\mu)/\mu = 1/\mu+1$.

{\bf SmallWorld networks}   We also studied symmetries, clusters, and subgroup decompositions in smallworld graphs. Smallworld graphs \cite{Watts1998, BollobasBOOK} were generated by starting with a ring of nearest neighbor connected nodes, then adding a fixed number of edges to give the same number of edges as the random graphs in the text. We found we had to add many edges beyond the usual few used to generate the smallworld effect because adding only a few edges beyond the ring rarely resulted in any symmetries.  As a result the smallworld examples approached being a like the random graphs so we do not display their results although the two systems each have symmetries that the other does not so they appear to not be exactly identical.

{\bf The Mesa del Sol power grid.}  In Supplementary Fig.~\ref{Suppfig:1} we show a circle plot of the Mesa del Sol network which, because of the network size (132 nodes), exposes the cluster structure much better. There are 20 nontrivial clusters and 10 subgroups in the decomposition. There are a large number of trivial clusters with only about 1/3 of the nodes being in a synchronizable cluster.  However, in that subset of clustered nodes the subgroup decomposition shows that ID is dynamically  possible.

\begin{figure}[tbp]
  \includegraphics[width=\textwidth]{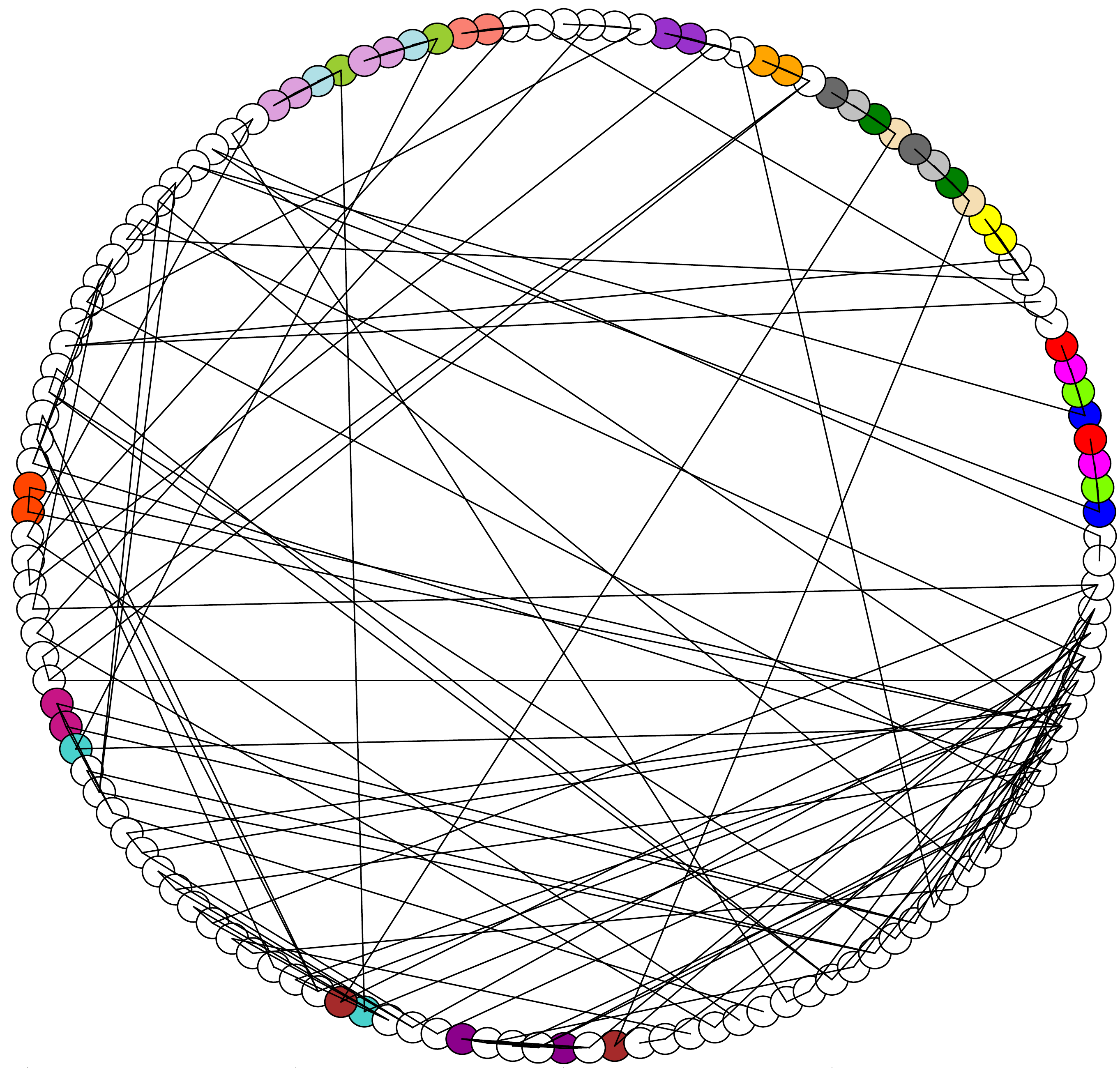}
  \caption{\textbf{Network and cluster structure of the Mesa del Sol electric grid.}  Colors are used to denote clusters.  Nodes colored white are trivial clusters, containing only one element.
  \label{Suppfig:1}}
\end{figure}

\end{document}